\documentclass[amsmath,amssymb,nofootinbib,aps,prc,two column ]{revtex4-1}
\usepackage{graphicx}
\usepackage{multirow}
\usepackage{epstopdf}
\usepackage{caption}
\usepackage{bm}
\usepackage{textcomp} 
\usepackage{subfigure}
\begin{document}

\title{The effect of hadronic scatterings on the measurement of vector meson spin alignments in heavy-ion collisions}
\author{Diyu Shen$^{a,b}$}
\author{Jinhui Chen$^{c}$} 
\author{Zi-Wei Lin$^{d,e}$} 

\affiliation{$^a$Shanghai Institute of Applied Physics, Chinese Academy of Sciences, Shanghai 201800, China}
\affiliation{$^b$University of Chinese Academy of Sciences, Beijing 100049, China}
\affiliation{$^c$Institute of Modern Physics and Key Laboratory of Nuclear Physics and Ion-beam Application (MOE), Fudan University, Shanghai 200433, China}
\affiliation{$^d$Key Laboratory of Quarks and Lepton Physics (MOE) and Institute of Particle Physics, Central China Normal University, Wuhan 430079, China}
\affiliation{$^e$Department of Physics, East Carolina University, Greenville, North Carolina 27858, USA}

\begin{abstract}
Spin alignments of vector mesons and hyperons in relativistic heavy-ion collisions have been proposed as signals of the global polarization. 
The STAR experiment first observed the $\rm \Lambda$ polarization. Recently, the ALICE collaboration measured the transverse momentum ($p_T$) and the collision centrality dependence of $K^*$ and $\phi$ spin alignments in Pb-Pb collisions at $\rm \sqrt{s_{NN}}$ = 2.76 TeV. A large signal is observed in the low $p_T$ region of mid-central collisions for $K^*$ while the signal is much smaller for $\phi$, and these have not been understood yet. Since vector mesons have different lifetimes and their decay products have different scattering cross sections, they suffer from different hadronic effects. In this paper, we study the effect of hadronic interactions on the spin alignment of $K^*$, $\phi$ and $\rho$ mesons in relativistic heavy-ion collisions with a multi-phase transport model. We find that hadronic scatterings lead to a deviation of the observed spin alignment matrix element $\rho_{00}$ away from the true value for $\rho$ and $K^*$ mesons (with a bigger effect on $\rho$) while the effect is negligible for the $\phi$ meson. The effect depends on the kinematic acceptance: the observed $\rho_{00}$ value is lower than the true value when the pseudorapidity ($\eta$) coverage is small while there is little effect when the $\eta$ coverage is big. Our study thus provides valuable information to understand the vector meson spin alignment signals observed in the experiments.
\end{abstract}

\maketitle

%%%%%%%%%%%%%%%%%%%%%%%%%%%%%%%%%%%%%%%%%%%%%%%%%%%%%%%%%
\section{Introduction} \label{sec:intro}	

It was predicted that a hot and dense matter, known as a quark-gluon plasma (QGP), will be formed in relativistic heavy-ion collisions~\cite{RevModPhys.89.035001}. This new state of matter~\cite{Gyulassy:2005,BRAUNMUNZINGER201676,CHEN20181} could have a large angular momentum with the direction  perpendicular to the reaction plane in non-central  collisions~\cite{PhysRevLett.94.102301,Voloshin:2004ha,PhysRevLett.120.012302}. The angular momentum is conserved during the evolution of the system and may result in spin-orbit coupling among the quarks, which will generate a net polarization of hyperon or vector mesons such as $\Lambda$, $\phi$, $K^*$ and $\rho$ due to the hadronization process~\cite{LIANG200520}. 
On the other hand, quarks and antiquarks may or may not be produced with an initial global polarization~\cite{PhysRevC.101.031901}. As the QGP approaches the transition to hadrons, the matter becomes strongly interacting due to nonperturbative effects and constitute quarks or antiquarks may approach spin and helicity equilibration with the vorticity~\cite{PhysRevC.101.031901}. 
Therefore, studying the spin polarization can provide additional dynamical information about the hot and dense matter. The topic is gaining increasing interests both in theory~\cite{PhysRevC.97.034917,PhysRevC.100.064904,PhysRevC.84.054910,PhysRevC.99.044910,Xia:2020tyd,Liang:2019pst} and experiments~\cite{Acharya:2019vpe,Acharya:2019ryw,Adam:2018ivw,STAR:2017ckg, PhysRevC.77.061902,ZHOU2019559}. More details can be found in some of the recent reviews~\cite{FLORKOWSKI2019103709,Liu:2020ymh,Gao:2020}. 

The spin alignment of a vector meson is described by a $3 \times 3$ Hermitian spin-density matrix with unit trace~\cite{PhysRevC.97.034917}. 
When the off-diagonal matrix elements are neglected or set to zero, 
the angular distribution of the decay products with respect to the system angular momentum in the vector meson's rest frame~\cite{SCHILLING1970397}
only depends on the 00-component of the matrix element ($\rho_{00}$) as
\begin{equation} \label{eq:1}
f(\theta^*) \equiv \frac{dN}{d(cos\theta^*)}=N_0 \times [(1-\rho_{00})+(3\rho_{00}-1)cos^2\theta^*].
\end{equation}
In the above, $N_0$ is a normalization factor, and $\theta^*$ is the angle between the decayed daughter and the system's orbit angular momentum $\hat{L}$ in the vector meson's rest frame. Experimentally, $\rho_{00}$ can be determined by measuring the angular distribution of Eq.\eqref{eq:1}. One sees that a deviation of the $\rho_{00}$ value from 1/3 means a spin alignment of the vector meson.

Measurements of vector meson spin alignments in heavy-ion collisions have been preformed. Results in Au+Au collisions at $\mathrm{\sqrt{s_{NN}}}$=200 GeV were initially consistent with $\rho_{00}$ = 1/3 within uncertainties~\cite{PhysRevC.77.061902} and then show a possible signal with improved event statistics~\cite{ZHOU2019559}. Data in Pb-Pb collisions at $\mathrm{\sqrt{s_{NN}}}$=2.76 TeV show that the $\rho_{00}$ of $K^*$ mesons could be significantly smaller than 1/3 at $p_T<1.0$ GeV/c while the  $\rho_{00}$ of $\phi$ mesons is much closer to 1/3~\cite{Acharya:2019vpe}. 
It is proposed that a strangeness current may exist in heavy-ion collisions and gives rise to a non-vanishing mean $\varphi$ field, which explains in part the $\phi$ meson spin alignment but does not apply to the $K^*$ meson~\cite{Sheng:2019}. The $\rho_{00}$ value could also depend on the quark hadronization process and the possible $p_T$ dependence of vorticity. So far the observed $\rho_{00}$ values of vector mesons have not been fully understood, especially in conjunction with the magnitude of the observed $\Lambda$ global polarization~\cite{Acharya:2019ryw,Adam:2018ivw,STAR:2017ckg,Huang:2020xyr}. 

Another physics process that may affect the observed $\rho_{00}$ values is hadronic scatterings of the decay products of vector mesons.
On the one hand, the decay daughters of $\phi$, $K^*$ and $\rho$ mesons are different and thus have different hadronic interactions. 
On the other hand, the lifetime of each vector meson species is different. 
One would expect that the vector meson with a longer lifetime will suffer less from hadronic interactions since their daughters are produced later at lower densities. Therefore we investigate quantitatively the effect of hadronic scatterings on vector meson spin alignments in Pb-Pb collision at $\mathrm{\sqrt{s_{NN}}}$ = 2.76 TeV with the string melting version of a multi-phase transport  (AMPT) model~\cite{PhysRevC.72.064901} (unless specified otherwise). The decay channels used in the current study are $\phi \rightarrow K^+ + K^-$, $K^{*}  \rightarrow K + \pi$ and $\rho \rightarrow \pi + \pi$ for the three vector meson species, respectively. 

The paper is organized as follows. The model and methodology are introduced in section~\ref{sec:methodology}. The results on the spin density matrix element $\rho_{00}$ of $\phi$, $K^*$ and $\rho$ before and after hadronic scatterings are presented in section~\ref{sec:final}. A summary is then given in Section~\ref{sec:summary}.

\section{Model and methodology} \label{sec:methodology}

\begin{figure}[htbp]
\vspace*{-0.1in}
	\includegraphics[scale=0.35]{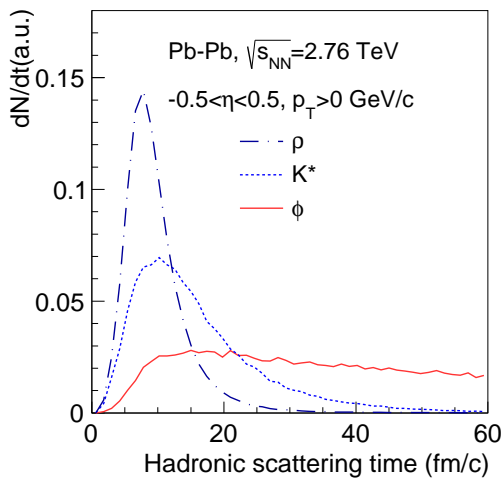}
	\captionof{figure}{The normalized decay time distribution of $\phi$, $K^*$ and $\rho$ mesons in the hadronic phase of Pb-Pb collisions at $\mathrm{\sqrt{s_{NN}}}$=2.76 TeV from the AMPT model. $\rho$ mesons decay quickly while $\phi$ mesons have a relative flat distribution after 10 fm/c.}
	\label{fig:lifetime}
\end{figure}

\begin{figure}
      \includegraphics[scale=0.55]{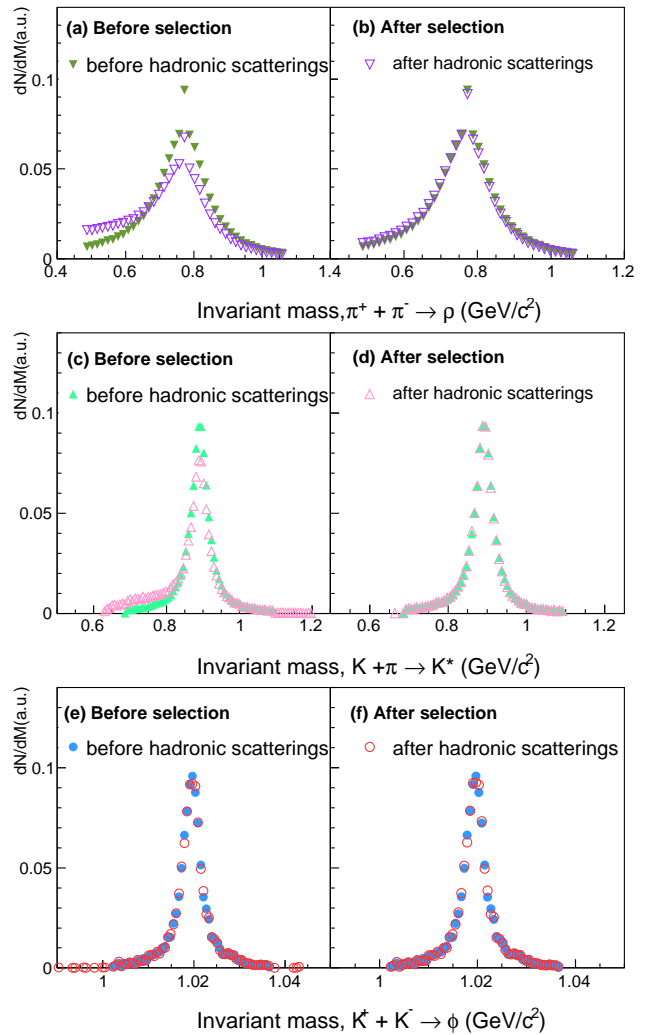}
      \captionof{figure}{The normalized invariant mass distribution of $\rho$, $K^*$, $\phi$ mesons from the AMPT model. In each panel the invariant mass is reconstructed by the decay channel as shown before (filled symbols) or after (open symbols) hadron scatterings. Open symbols in panels (b), (d) and (f) represent the distributions after removing those resonances that have a decay daughter with momentum change more than $0.01$ GeV/$c$ due to hadronic scatterings.}
	\label{fig:InvMass}
\end{figure} 

The AMPT model is a multi-phase transport model~\cite{PhysRevC.72.064901} for studying heavy-ion collisions. 
In this model, the initial conditions are taken from the spatial and momentum distributions of minijet partons and soft string excitations from the HIJING event generator~\cite{PhysRevD.44.3501}, which is followed by two-body elastic parton scatterings using the parton cascade model ZPC~\cite{ZHANG1998193}; the conversion from partons to hadrons via either the string fragmentation~\cite{SJOSTRAND199474} 
for the default version or a quark coalescence model for the string melting version 
~\cite{Lin:2001zk,He:2017tla}, and hadronic scatterings based on an extended relativistic transport model ART~\cite{PhysRevC.52.2037}.
Since the spin degree of freedom is not considered in the current AMPT model, to simulate a spin alignment signal we redistribute the decay products according to Eq.\eqref{eq:1} when resonances decay~\cite{LAN2018319} and then study the hadronic scattering effect on vector meson spin alignment with different input $\rho_{00}$ values in this work. For hadronic interactions, the extended ART model includes baryon-baryon, baryon-meson, and meson-meson elastic and inelastic scatterings~\cite{PhysRevC.72.064901}. In general, the cross sections of elastic or inelastic scatterings depend on the center of mass energy of the scattered hadrons. Therefore hadrons at different momentum have different cross sections, which may contribute to the $p_T$ dependence of the observed $\rho_{00}$ value of vector mesons. 
As the global vorticity is expected to peak at semi-central collisions~\cite{PhysRevLett.94.102301}, we choose the impact parameter $b$=8 fm to mimic such collisions in this study. 

A decay daughter that has a large transverse momentum is more likely to come from a parent hadron with a large transverse momentum. When one measures the $p_T$ dependence of vector meson spin alignment, it will be influenced by the $p_T$ dependent hadronic interactions. In addition, the decay times of $\phi$, $K^*$ and $\rho$ mesons are different as illustrated in Fig.~\ref{fig:lifetime}. The $\rho$ meson has the shortest lifetime, therefore we expect it to be affected most strongly by the hadronic scatterings. However, the $\phi$ meson is expected to be less affected due to its long lifetime. Furthermore, pions as the decay products of $\rho$ meson are expected to be scattered more frequently than kaons as the decay products of $\phi$ meson. Since the $K^*$ meson has a moderate lifetime with kaon and pion as the decay products, one would expect it to be moderately affected by hadronic scatterings. Therefore it is worthwhile to quantify how much hadron scatterings could affect the final spin alignment results of different vector mesons.

In experiments one uses the $n$-th event plane ($n=1, 2$) to estimate the direction of the system's orbital angular momentum $\hat{L}$. In this paper, we calculate the direction of $\hat{L}$ with participant nucleons in each event.
The decay daughters from the aforementioned channels of interest are labelled so that we know explicitly which two decay daughters have come from the same parent hadron. 
The phase space information of the decay daughters is also recorded when the decay happens, which enables us to distinguish whether a decay product has experienced scatterings or not by comparing the momentum information upon decay with that after the hadron cascade. When a decay daughter is destroyed due to a subsequent inelastic scattering, naturally the parent meson cannot be reconstructed from the final hadron record. 
In addition, experiments usually reconstruct non-stable vector mesons via the invariant mass distribution of the candidate decay daughters, 
where the invariant mass window is determined by the vector meson physical width in convolution with the detector momentum resolutions~\cite{Acharya:2019vpe,PhysRevC.77.061902}.
As a result, a parent meson with any decay daughter having subsequent elastic scattering(s) will likely not be reconstructed with the experimental method of background subtraction within the invariant mass window. Therefore, in the selection procedure of this study we remove a resonance if it has a decay daughter with a momentum change more than $0.01$ GeV/c due to elastic scattering(s).
 
Figure~\ref{fig:InvMass} shows the normalized invariant mass distributions of the vector mesons reconstructed via their decay products before and after the selection procedure as described above. It shows in panels (a), (c) and (e) that the hadronic scatterings change the shape of the reconstructed invariant mass distributions. The degrees of change are different for $\phi$, $K^*$ and $\rho$ mesons, where the $\rho$ meson distribution changes most significantly as expected while the $\phi$ meson distribution has little change. After the selection procedure where vector mesons with scattered decay daughters are removed, the distributions after hadronic scatterings, as shown in panels (b), (d) and (f), are similar to those before hadronic scatterings. The decay daughters that have passed the selection procedure are then used in our analysis of the vector meson spin alignment. Note that the distributions of Fig.~\ref{fig:InvMass} with different momentum difference cuts have been checked and similar results are obtained.

\section{Results and discussion} \label{sec:final}

It has been pointed out that a finite acceptance in experiments will lead to an increase of the observed $\rho_{00}$ as the range of pseudorapidity $\rm -\eta_{max} < \eta < \rm \eta_{max}$ is small ~\cite{LAN2018319}, where $\rm \eta_{max}$ denotes the maximum $|\eta|$ of decay daughters. To focus on the hadronic scattering effect, we first correct for the acceptance effect. 
When the effect of hadron scatterings is neglected, we can write the decay daughter distribution in the vector meson rest frame within a specific kinematic window as
\begin{eqnarray} \label{eq:2}
\rm f_{obs}(\theta^*, \rho_{00}^{in}, \eta_{max}, p_T)&=& \rm f_{true}(\theta^*, \rho_{00}^{in})\times \nonumber\\
\rm g(\theta^*, \eta_{max}, p_T),
\end{eqnarray}
where $\rm g(\theta^*, \eta_{max}, p_T)$ denotes the effect of the kinematic window,
$\rho_{00}^{in}$ is the input $\rho_{00}$ value, $\rm f_{obs}(\theta^*, \rho_{00}^{in}, \eta_{max},p_T)$ is the observed distribution, and $\rm f_{true}(\theta^*, \rho_{00}^{in})$ is the true distribution for a given input $\rho_{00}$.
Note that in the above we assume that the effect of finite acceptance (i.e., the $g$ function) is independent of the input $\rho_{00}$ value, which we have verified numerically to be true to a good accuracy. 
Since the $\rm f_{true}(\theta^*, \rho_{00}^{in})$ distribution is flat for $\rho_{00}^{in}=1/3$, the nontrivial distribution before hadronic scattering represents the effect from the kinematic window and can be used as the correction function:a
\begin{eqnarray} \label{eq:3}
\rm \frac {f_{obs}(\theta^*, \rho_{00}^{in}, \eta_{max}, p_T)} {f_{obs}(\theta^*, \rho_{00}^{in}=1/3, \eta_{max}, p_T)} &=& \rm \frac {f_{true}(\theta^*, \rho_{00}^{in})} {f_{true}(\theta^*, \rho_{00}^{in}=1/3)} \nonumber\\
=\rm A ~ f_{true}(\theta^*, \rho_{00}^{in}), 
\end{eqnarray}
where $A$ is a constant. Therefore, for each resonance we divide the $\rm cos(\theta^*)$ distribution of interest by the $\rm cos(\theta^*)$ distribution obtained for the $\rho_{00}$=1/3 case without hadronic scatterings with the same kinematic window. The spin alignment signal is then calculated from a fit to the corrected distribution.

\begin{figure}[htbp]
\vspace*{-0.01in}
	\includegraphics[scale=0.4]{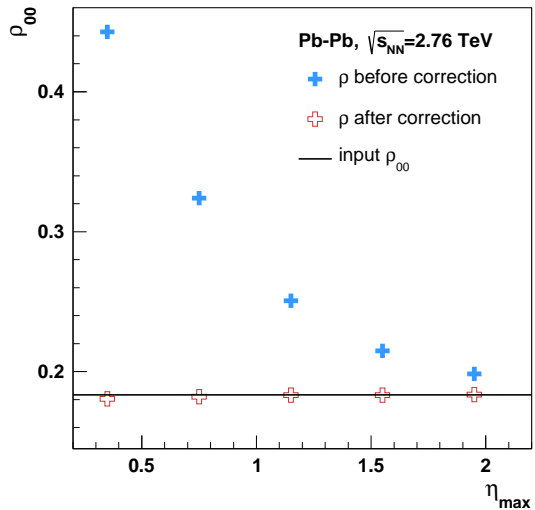}
	\captionof{figure}{Correction of the $\eta$ window dependence of $\rho_{00}$ for $\rho$ meson in Pb-Pb collisions at $\mathrm{\sqrt{s_{NN}}}$ = 2.76 TeV (without hadronic scatterings and with full $p_T$ range), where $\rm \eta_{max}$ represents the maximum $|\eta|$ of decay daughters. The filled and open symbols represent the results without and with the acceptance correction, respectively; and the solid line represents the input value $\rho_{00}=0.183$.}
	\label{fig:fiteg}
\end{figure}

Figure~\ref{fig:fiteg} shows an example of the acceptance correction. The filled symbols represent the extracted $\rho_{00}$ values of the $\rho$ meson without acceptance correction (and without hadronic scatterings), where the magnitude increases with the decrease of $\rm \eta_{max}$~\cite{LAN2018319}. Our results after the acceptance correction, shown as the open symbols, are consistent with the input $\rho_{00}$ value (solid line).
There is no $\eta$ cut on the vector mesons. We have also checked the acceptance correction with the method in Ref.~\cite{PhysRevC.98.044907}, which is usually applied in the experimental analysis, and obtained consistent results.

\begin{figure*}
	\centering
	{\includegraphics[width=0.9\linewidth]{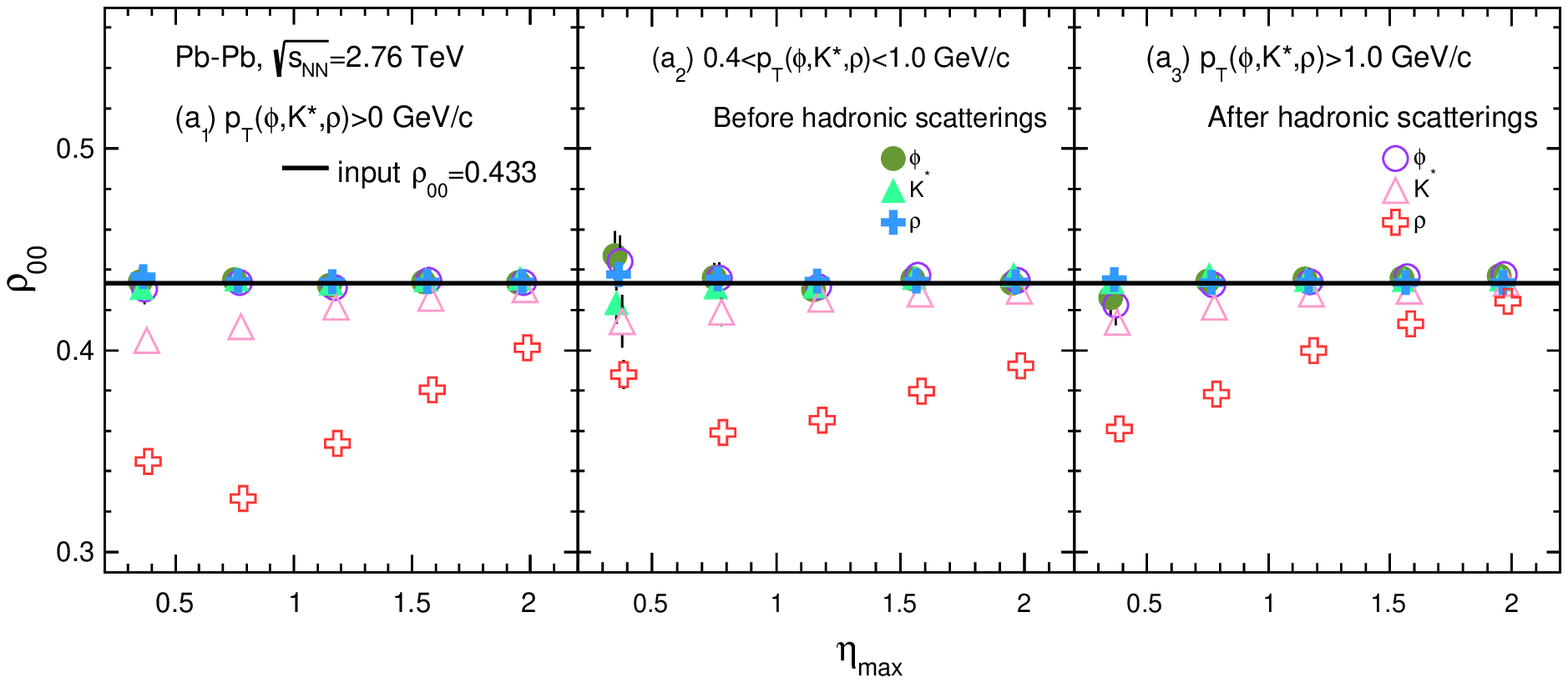}} \\
	{\includegraphics[width=0.9\linewidth]{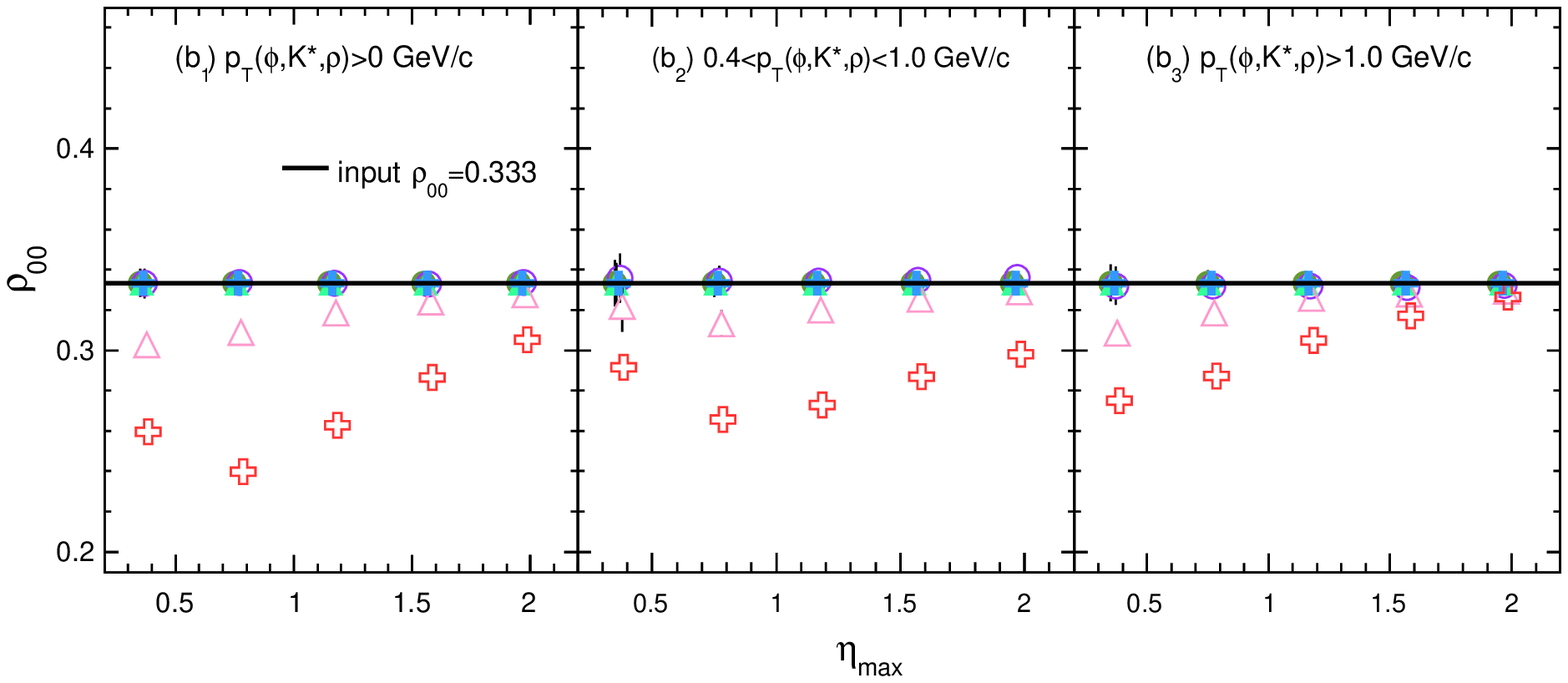}}\\
    {\includegraphics[width=0.9\linewidth]{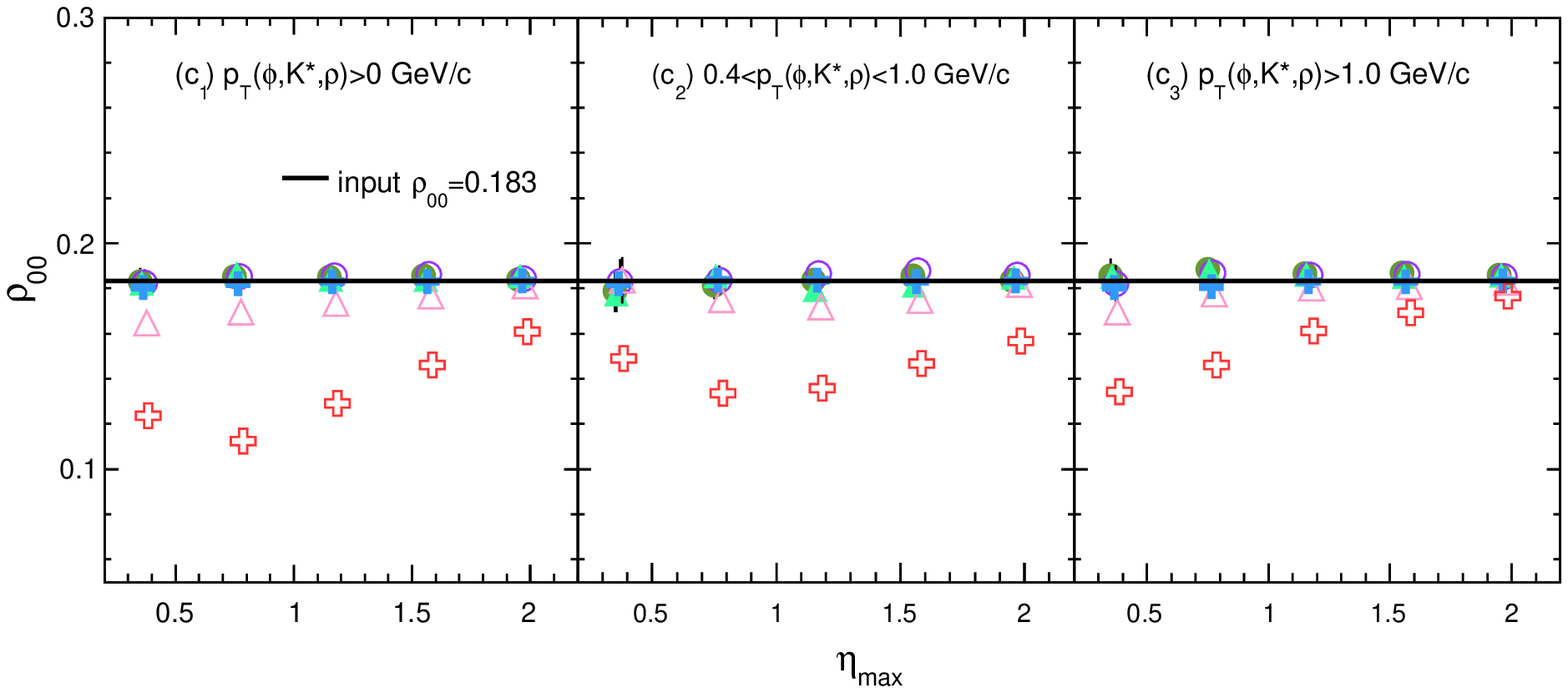}}\\
    \captionof{figure}{The extracted $\rho_{00}$ values of $\phi$, $K^*$ and $\rho$ mesons as functions of the $\eta$ range in Pb-Pb collisions at $\mathrm{\sqrt{s_{NN}}}$=2.76 TeV and $b=8$ fm from the AMPT model with three different input values of $\rho_{00}$ without and with hadronic scatterings. The x axis center of each point has been slightly shifted to distinguish their error bars.}
	\label{fig:PbPb276}
\end{figure*}

In order to investigate the possible $p_T$ dependence of the hadronic scattering effect, the vector mesons $\phi$, $K^*$ and $\rho$ have been classified into three $p_T$ intervals: the full $p_T$ range, 0.4 $<p_T<$ 1 GeV/c, and $p_T>$ 1 GeV/c, similar to the $p_T$ binning applied in experiments~\cite{Acharya:2019vpe,ZHOU2019559}. Meanwhile, the decay daughters of each resonance species have been selected within different $\eta$ ranges ($\rm \eta_{max}=$ from 0.4 to 2) to study the hadronic scattering effect under different experimental acceptance. Figure~\ref{fig:PbPb276} presents the $\rho_{00}$ values of $\phi$, $K^*$ and $\rho$ mesons as functions of $\rm \eta_{max}$ in Pb-Pb collisions at $\mathrm{\sqrt{s_{NN}}}$=2.76 TeV with the input $\rho_{00}$ values of 0.433, 0.333 and 0.183. The global cutoff time for the hadron cascade in the AMPT model is set to 60 fm/$c$. The three sub-panels of (a), (b) or (c) show the extracted $\rho_{00}$ in the three $p_T$  ranges, respectively, where filled symbols represent the results without hadronic scatterings and open symbols represent the results after hadronic scatterings. While the $\rho_{00}$ value without hadronic scatterings (after the correction for finite acceptance) is consistent with the input value for most cases, sizable effects from hadronic scatterings can be seen for the $\rho$ meson. The $\rho_{00}$ values of $K^*$ and $\rho$ mesons after hadronic scatterings are found to decrease and the decrease is typically bigger for a smaller $\eta$ range (except for some cases when $\rm  \eta_{max}<0.8$). The $\rho$ meson suffers from the most significant hadronic effect as expected, while the $\phi$ meson is basically not affected mainly due to its long lifetime.

One also sees that the change of $\rho_{00}$ due to hadronic scatterings is quite similar for the three different input $\rho_{00}$ values and three $p_T$ intervals. 
The hadronic scattering effect on $\rho$ mesons is stronger than that on $K^*$ mesons as expected, while the effect on $\phi$ mesons is negligible in all $p_T$ intervals.
The general decrease of $\rho_{00}$ may be understood due to the anisotropy of the scattering probabilities, where the decay daughters at $\theta^* \sim 0$ or $\pi$ (which go along the $\pm y$ directions for a parent hadron at rest) are more likely to be scattered than those at $\theta^* \sim 90^{\circ}$ (which go inside the $x-z$ plane for a parent hadron at rest). 
We also see that for $\rho$ mesons the decrease of $\rho_{00}$ for input $\rho_{00}$=0.433 is slightly stronger than that for input $\rho_{00}$=0.183, especially at small $\rm \eta_{max}$. 
This may be due the fact that there are more decay daughters emitted along $\theta^* \sim 0$ or $\pi$, which further increases the scattering probabilities along these directions.

\begin{figure*}
	\centering
	{\includegraphics[width=0.9\linewidth]{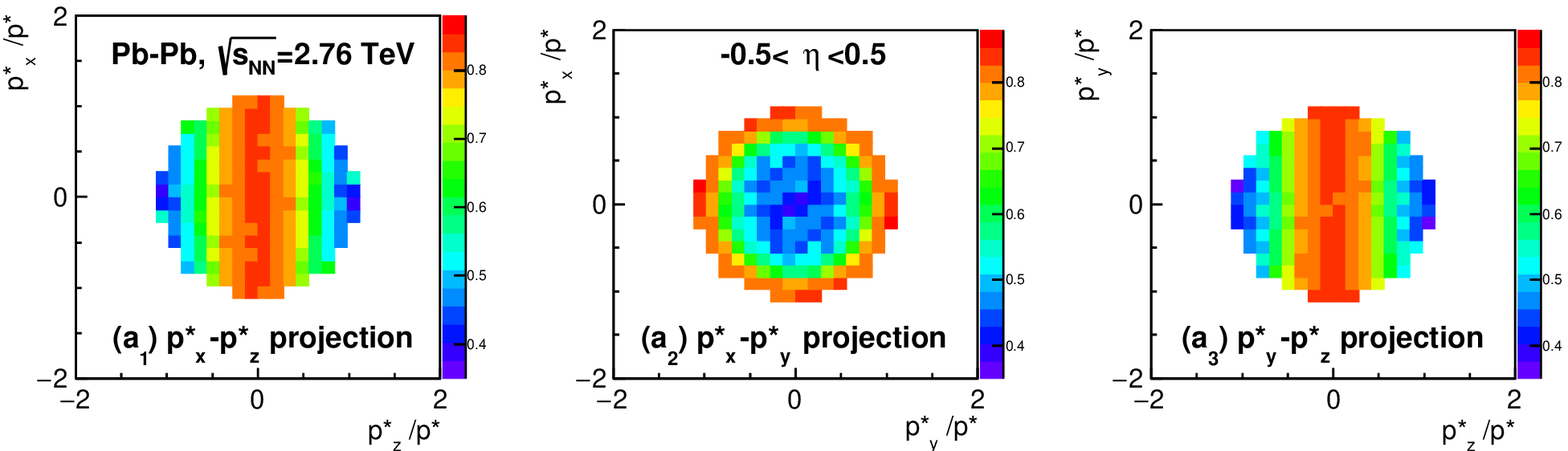}} \\
	{\includegraphics[width=0.9\linewidth]{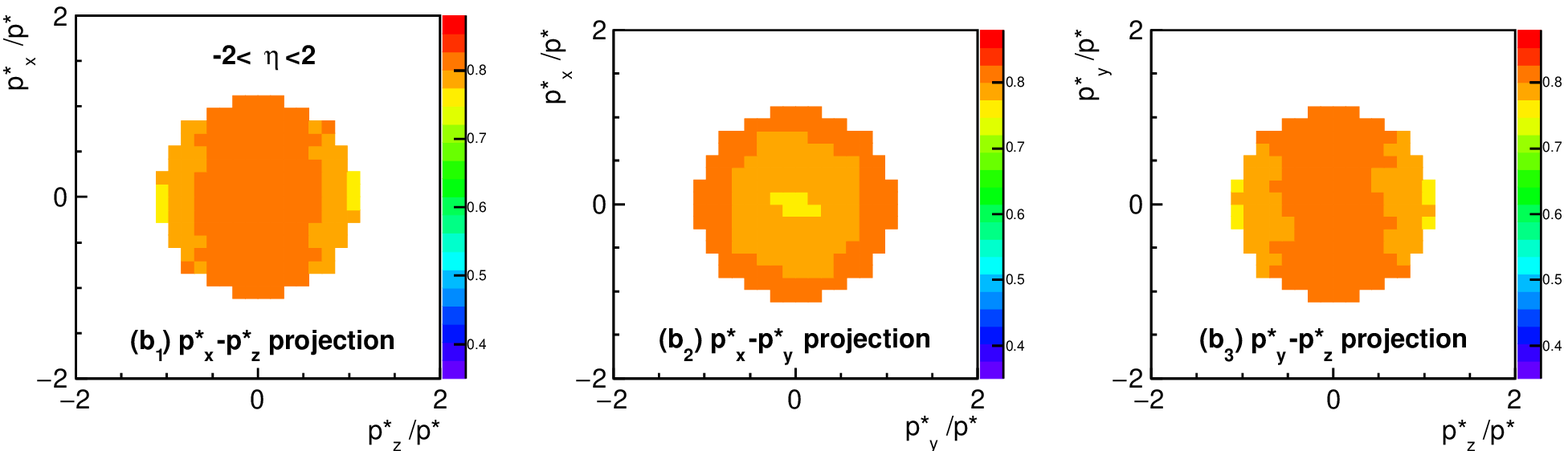}} \\
	\captionof{figure}{The probability distribution of $\rho$ mesons lost due to hadronic scatterings in Pb-Pb collisions at $\mathrm{\sqrt{s_{NN}}}$=2.76 TeV with input $\rho_{00}=0.333$; $p^*$ represents the decay daughter's momentum in the $\rho$ meson rest frame.}
	\label{fig:lost-Dau}
\end{figure*}

\begin{figure*}
	\centering
	{\includegraphics[width=0.9\linewidth]{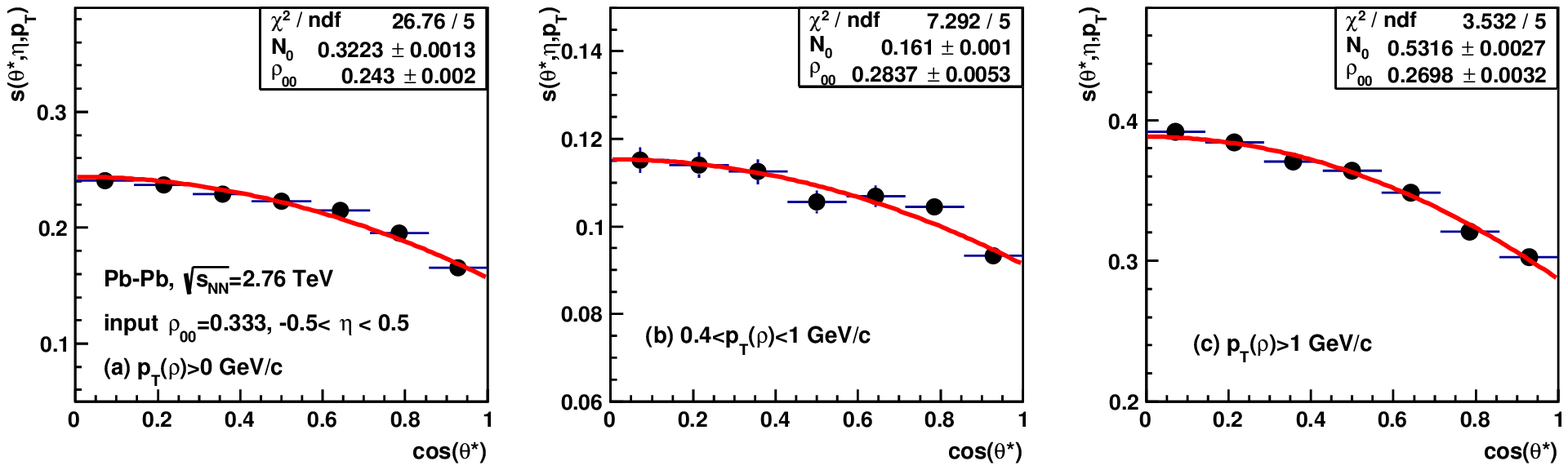}}
	\captionsetup{justification=raggedright}
	\captionof{figure}{The effect of hadronic scatterings as a function of $\cos(\theta^*)$ in Pb-Pb collisions at $\rm \sqrt{s_{NN}}=$ 2.76 TeV with input $\rho_{00}=0.333$ and $\rm \eta_{max}$=0.5.}
	\label{fig:ratio}
\end{figure*}

The probability of decay daughters to be scattered, and thus the influence of hadronic scatterings on the vector meson spin alignment, is not isotropic but depends on the effective three-dimensional geometry of hadronic matter. Including the hadronic effect, we may write the observed $\rm cos(\theta^*)$ distribution approximately as
\begin{eqnarray} \label{eq:4}
  \rm f'_{obs}(\theta^*, \rho_{00}^{in}, \eta_{max},p_T) &=& \rm  f_{obs}(\theta^*, \rho_{00}^{in}, \eta_{max},p_T) \times \nonumber\\
  s(\theta^*, \rho_{00}^{in}, \eta_{max},  p_T),
\end{eqnarray}
where $\rm f'_{obs}(\theta^*, \rho_{00}^{in}, \eta_{max},p_T)$ is defined as the distribution after hadronic scattering, $\rm f_{obs}(\theta^*, \rho_{00}^{in}, \eta_{max},p_T)$ is the one before hadronic scattering, and $\rm s(\theta^*, \rho_{00}^{in}, \eta_{max}, p_T)$ represents the fraction of vector mesons that survive the hadronic scattering. Therefore, the probability of losing a vector meson due to hadronic scattering (i.e., having a vector meson with scattered daughters) can be written as
\begin{eqnarray}\label{eq:5}
\rm 1-s(\theta^*, \rho_{00}^{in}, \eta_{max}, p_T)
=1-\frac{\rm f'_{obs}(\theta^*, \rho_{00}^{in}, \eta_{max},p_T)} {\rm f_{obs}(\theta^*, \rho_{00}^{in}, \eta_{max},p_T)}.  
\end{eqnarray}
Figure~\ref{fig:lost-Dau} presents the above probability distributions for $\rho$ mesons after the projection to two-dimensional relative momentum planes. Variables $p_x^*$ and $p_z^*$ represent the decay daughter's momentum in the parent's rest frame along the impact parameter and the beam direction, respectively. The momentum is normalized with  $p^*=\sqrt{{p_x^*}^2+{p_y^*}^2+{p_z^*}^2}$ because the distribution of $p_y^*/p^*=\cos \theta^*$ is directly related to the spin alignment and the $\rho_{00}$ value. We see that the decay daughters of $\rho$ mesons are more likely to be scattered along the $p_y^*$ axis (also the $p_x^*$ axis) than along the $p_z^*$ axis, which leads to a bigger suppression at $\theta^* \sim 0$ or $\pi$ than at $\theta^* \sim 90^{\circ}$ and thus a decrease of the extracted $\rho_{00}$ value. By comparing the distributions for $\rm \eta_{max}=$0.5 and those for $\rm \eta_{max}=$2, one observes that the anisotropy of the scattering probability is bigger for the narrower $\eta$ range, which corresponds to a stronger decrease of the extracted $\rho_{00}$ value.

The observed probability distributions can be understood in terms of the effective geometry of the matter. In the case of $\rm \eta_{max}$=0.5, the effective geometry of the hadronic matter as seen by the decay daughters in the parent's rest frame may be considered as a short cylinder with the axis along the $z$ axis. Therefore scatterings of decay daughters in the transverse plane are more likely than those along the $z$ axis, as illustrated in the Fig.~\ref{fig:lost-Dau}. In particular, as shown by the projection in the $p_y^*$-$p_z^*$ plane, scatterings along the $p_z^*$ axis are less likely, which leads to relatively more vector mesons to be observed around $\rm \theta^* \sim 90^\circ$ and a smaller observed $\rho_{00}$ value. On the other hand, from panels ($a_2$) and ($b_2$) of Fig.~\ref{fig:lost-Dau} one sees that the scattering probability is about symmetric along $p_x^*$ and $p_y^*$ axes, indicating that the transverse spatial anisotropy in the hadronic stage (or its effect on the scattering probability versus $\cos(\theta^*)$) is small.

Figure~\ref{fig:ratio} shows the surviving function $\rm s(\theta^*, \rho_{00}^{in}, \eta_{max},  p_T)$ of $\rho$ mesons in the case of input $\rho_{00}$=0.333 $\rm~and~\eta_{max}$=0.5. We see that a $\rho$ meson at higher $p_T$ is more likely to survive the hadronic scatterings. A non-uniform distribution of $\rm s(\theta^*, \rho_{00}^{in}, \eta_{max}, p_T)$ indicates the hadronic effect, and the decreasing trend versus $\cos \theta^*$ corresponds to a decrease of the extracted $\rho_{00}$ value. Also note that the $\cos \theta^*$ distribution after hadronic scatterings and the finite acceptance correction may not follow the shape of Eq.\eqref{eq:1} well; this can be reflected in the sometimes large $\chi^2$ values in the fit with Eq.\eqref{eq:1}, as shown in Fig.~\ref{fig:ratio}(a) as an example.

We have also calculated Au+Au collisions at $\mathrm{\sqrt{s_{NN}}}$=200 GeV and the hadronic effect there is found to be smaller than that in Pb-Pb collisions at $\mathrm{\sqrt{s_{NN}}}$=2.76 TeV. This could come from the lower hadron multiplicity and consequently the smaller scattering probability of decay daughters in the hadron cascade of the lower-energy Au+Au collisions.

\begin{figure}
\centering
{\includegraphics[width=0.9\linewidth]{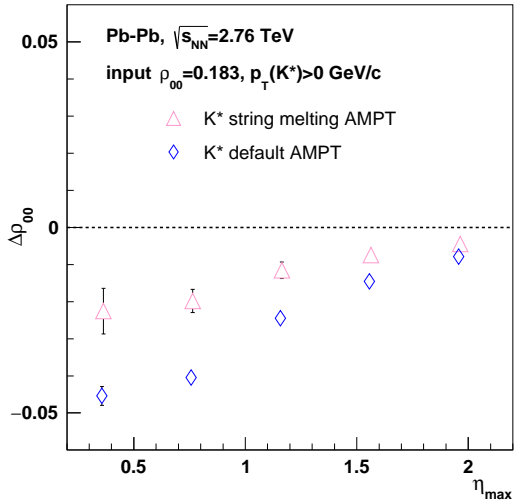}}
\captionof{figure}{
The change of the extracted $\rho_{00}$ values of $K^*$ mesons due to hadronic scatterings from the string melting version (triangles) and  the default version (diamonds) of the AMPT model for Pb-Pb collisions at 
$\mathrm{\sqrt{s_{NN}}}$=2.76 TeV.}
\label{fig:com}
\end{figure}

In principle, the hadronic effect on vector meson spin alignments shall be present in any hadronic transport model~\cite{Bratkovskaya:2011wp,Weil:2016zrk},  
although the magnitude of the effect could be different 
due to the different treatments of the hadron transport. 
For example, the string melting version of the AMPT model 
starts the quark coalescence process after the kinetic freezeout of partons.  
Therefore, the magnitude of the parton cross section affects the average density at the start of hadron cascade ~\cite{Zhang:2008zzk} and may thus affect the magnitude of the effect of hadronic scatterings on spin alignment observables. 
Here we check the effect in the AMPT model with a different configuration. Figure~\ref{fig:com} presents the results from the default version of AMPT, where the parton cascade only includes minijet partons and is thus shorter and the hadronization is modeled by the Lund string fragmentation~\cite{SJOSTRAND199474}. Since the $\rho$ meson global spin alignment has not been measured yet, we choose the $K^*$ for illustration.
We see that the general feature is the same, while the decrease of the extracted $\rho_{00}$ in the default AMPT model is larger, which is a result of the earlier start of the hadron cascade phase in the default AMPT model. Future experimental measurement on the $\rho$ meson spin alignment is called for.

Note that this study shows that hadronic scatterings could lead to a finite deviation of the extracted $\rho_{00}$ value from the true value at the time of the vector meson decays. We have not addressed what the true value shall be or how the true value is developed. One may expect that vector mesons with a longer lifetime would experience more hadronic scatterings and their later freeze-out time could affect their polarization including the $\rho_{00}$ value. However, an explicit calculation would require the inclusion of the spin degree of freedom in the partonic and hadronic transport, which is beyond the scope of the current study.

\section{Summary} \label{sec:summary}
We have studied the effect of hadronic scattering on the spin alignment of $\phi$, $K^*$ and $\rho$ mesons at LHC energies with a multi-phase transport model. We find that hadronic interactions will lead to a deviation of the extracted spin density matrix element $\rho_{00}$ from its true value, where the deviation depends on the effective three-dimensional geometry of the hadronic matter. With finite acceptance, the observed $\rho_{00}$ decreases due to hadronic scattering because it is less likely for the decay daughters to be scattered along the $z$ axis, which tends to around $\theta^* \sim 90^\circ$. The hadronic effect on $\rho_{00}$ is more significant for $\rho$ mesons
because of their shorter lifetime and the larger scattering cross section of the $\pi$ decay daughters, therefore measurements of the $\rho$ meson spin alignment will be interesting. The hadronic effect on $K^*$ mesons is moderate, while $\phi$ mesons are almost not affected by hadronic scatterings due to their longer lifetime and the relatively small scattering cross sections of kaons.
Furthermore, the hadronic effect on $\rho_{00}$ is bigger from the default version of the AMPT model than that from the string melting version because of the earlier and denser hadron matter in the default AMPT model. 
Our study suggests that hadronic scatterings could affect vector meson spin alignment observables in addition to the spin-orbit coupling and vorticity.

\section*{Acknowledgements}
The authors thank Shi Shusu for helpful discussions. The work of D. Y. Shen and J. H. Chen was supported in part by the Guangdong Major Project of Basic and Applied Basic Research No. 2020B0301030008,
the Strategic Priority Research Program of Chinese Academy of Sciences with Grant No. XDB34030000, and the National Natural Science Foundation of China under Contract Nos. 12025501, 11890710, 11890714 and 11775288. 
\bibliography{ref}

\end{document}